\documentclass{article}
\newcommand{\bfr}{\begin{flushright}}
\newcommand{\efr}{\end{flushright}}
 
\begin{document}
\title{Three-dimensional black holes and solitons in higher-dimensional theories with
compactification
}
\author{ 
Takuya Maki\\
Institute of Physics, Kitasato University,\\
Kitasato, Sagamihara-shi,
Kanagawa 228, Japan\\
and\\
Kiyoshi Shiraishi\\
Akita Junior College, \\
Shimokitade-sakura, Akita-shi,
Akita 010, Japan
}
\date{Class. Quantum Grav. {\bf 11} (1994) pp. 2781--2787
}
\maketitle
\begin{abstract}
Several types of static solutions to Einstein's equations coupled with antisymmetric tensor
fields are found in $(2+N+1)$-dimensional spacetime. The solutions describe a product of a
three-dimensional radially symmetric spacetime and an internal maximally symmetric
manifold. The scale of the internal space may depend on the radial distance from the origin
in these solutions. 
\end{abstract}

\bigskip

Recently the study of gravity in $(2+1)$ dimensions has attracted much attention. 
The discovery of the three-dimensional black hole solution by Ba\~nados et al.~(BTZ)
\cite{1} was a little surprising for theoretical physicists. Since such a non-trivial
geometry was found, the structure of the geometry, the relation to string theory, and the
behaviour of quantized fields in spacetime have been investigated by many authors \cite{2}.

More recently, Chan and Mann found a family of static soliton solutions in $(2+ 1)$-dimensional 
gravity coupled to a dilaton with an exponential potential term \cite{3}. On the other
hand, it is well known that certain types of dilaton fields naturally arise from dimensional
reduction of higher-dimensional theories \cite{4} (and references therein)%
\footnote{For higher-dimensional theories, see the reviews \cite{5}.}.

In the present paper we consider $(2+N+1)$-dimensional gravity coupled to
antisymmetric gauge fields and explicitly derive static soliton solutions in two-dimensional
space with $N$-dimensional compact internal manifold. We assume the scale of the internal
manifold may depend on the radial coordinate in two-dimensional space. This type of
solution can be regarded as an analogue of `local compactification' \cite{6}.

\bigskip

We shall deal with the Einstein gravity with antisymmetric tensor fields in $(2+N+1)$
dimensions. Further, we assume that there is no cosmological constant in $(2+N+1)$-dimensional
spacetime. The action (where we omit boundary terms) we consider is%
\footnote{Owing to a duality, a $3$-form field strength may be adopted instead of the $N$-form
one, to obtain the same solutions for the metric.}
\begin{equation}
S=\int d^{2+N+1}x \sqrt{-g}\left[\frac{1}{2\kappa^2}R-\frac{1}{4}F^2_{MN}-
\frac{1}{2N}F^2_{(N)}\right]\,,
\label{(1)}
\end{equation}
where $F_{MN}$ is a Maxwell field strength, while $F_{(N)}$ is an $N$-form gauge field strength.

The convention for indices is as follows. The $(2+N+1)$-dimensional indices are
denoted by $M, N,\ldots$. They are decomposed into $(2+1)$-dimensional indices
$\mu, \nu,\ldots$ and $N$-dimensional indices $m, n,\ldots$.

The field equations are easily seen to be
\begin{eqnarray}
& &R_{MN}-\frac{1}{2}R g_{MN}=\kappa^2T_{MN}\,,\label{(2)}\\
& &\partial_M(\sqrt{-g}F^{MN})=0\,,\label{(3)}\\
& &\partial_M(\sqrt{-g}F^{MN_1\ldots N_N}_{(N)})=0\,,\label{(4)}\\
& &T_{MN}=F^2_{MN}-\frac{1}{4}F^2g_{MN}+F^2_{(N)MN}-\frac{1}{2N}F^2_{(N)}g_{MN}\,,
\label{(5)}
\end{eqnarray}
where $F^2_{MN}=F_{PM}F^P{}_N$, $F^2_{(N)MN}=F_{(N)P_1\ldots P_{N-1}M}F^{P_1\ldots
P_{N-1}}_{(N)}{}_N$,
$F^2=F_{PQ}F^{PQ}$, and $F^2_{(N)}=F_{(N)P_1\ldots P_{N}}F^{P_1\ldots P_{N}}_{(N)}$.

We assume that the metric takes a block diagonal form in three- and $N$-dimensional
space. The internal space is assumed to be homogeneous and to admit an Einstein metric,
and that only the overall factor of the space may depend on the coordinates in three dimensions.

We wish to find the static solution to the equations (\ref{(2)})-(\ref{(5)}) which can be
interpreted as a product of a three-dimensional radially symmetric soliton and an
$N$-dimensional internal space whose scale varies with the radial coordinate of the tree
dimensions. The metric can be written as the sum of a three-dimensional rotationally
invariant static metric and  an $N$-dimensional maximally symmetric metric:
\begin{equation}
ds^2=e^{2N\phi(r)}\left[-\Delta(r)dt^2+\frac{dr^2}{\Delta(r)}+R(r)^2d\theta^2\right]
+b^2e^{-2\phi(r)}\tilde{g}_{mn}dx^m dx^n\,,
\label{(6)}
\end{equation}
where $\tilde{g}_{mn}$ is the metric of the $N$-dimensional extra space. The constant $b$ scales the internal
space. We assume that maximal symmetry of the extra space and its Ricci tensor can be expressed as
\begin{equation}
\tilde{R}_{mn}=k\tilde{g}_{mn}\,,
\label{(7)}
\end{equation}
where $k$ is a constant, and be rescaled to be $1, 0$ or $-1$.

For an electric point charge, the electric field is obtained by solving equation (\ref{(3)})
as
\begin{equation}
F_{tr}=e^{2N\phi(r)}\frac{Q}{R(r)}\,,
\label{(8)}
\end{equation}
where $Q$ is a constant which may be called an `electric charge'.

Equation (\ref{(4)}) can be solved by the Freund-Rubin ansatz \cite{7}, for which all
components of the antisymmetric tensor field $F_{M\ldots N}$ vanish except
\begin{equation}
F_{(N)m_1\ldots m_N}=q\epsilon_{(N)m_1\ldots m_N}\,,
\label{(9)}
\end{equation}
where $q$ is a constant.

Substituting the metric ansatz (\ref{(6)}) and equations (\ref{(7)})-(\ref{(9)}) into the
field equation (\ref{(2)}) with (\ref{(5)}), we obtain simultaneous differential equations
for
$\Delta(r)$,
$R(r)$, and
$\phi(r)$. After some manipulation, the equations are reduced to
\begin{eqnarray}
& &e^{-2N\phi}\Delta\left[\frac{R''}{R}+N(N+1)(\phi')^2\right]=0\,,\label{(10)}\\
& &\frac{N(N-1)}{b^2}k e^{2\phi}-e^{-2N\phi}\Delta\left[\frac{R''}{R}+
\frac{\Delta'}{\Delta}\frac{R'}{R}\right]\nonumber \\
& &\quad=\kappa^2\left[\frac{Q^2}{R^2}+(N-1)!\frac{q^2}{b^{2N}}e^{2N\phi}\right]\,,
\label{(11)}\\
& &\frac{N(N-1)}{b^2}k e^{2\phi}-e^{-2N\phi}\Delta\left[\phi''+
\left(\frac{\Delta'}{\Delta}+\frac{R'}{R}\right)\phi'\right]\nonumber \\
&
&\quad=\frac{\kappa^2}{N+1}\left[\frac{Q^2}{R^2}+(N-1)!\frac{2q^2}{b^{2N}}e^{2N\phi}\right]\,,
\label{(12)}
\end{eqnarray}
where the prime denotes the derivative with respect to $r$.

The equation (\ref{(10)}), which is a combination of the $tt$ and $rr$ components of the
Einstein equation, can be solved by assuming the following simple $r$-dependence for $R$
and $\phi$, also adopted by Chan and Mann \cite{3}:
\begin{equation}
R(r)=\gamma r^{\alpha}\quad\mbox{and}\quad \phi(r)=\beta\ln\frac{r}{r_0}\,,
\label{(13)}
\end{equation}
where $\alpha$, $\beta$, $\gamma$, and $r_0$ are constants. Taking these ansatze, we can see a relation between $\alpha$
and  $\beta$ from equation (\ref{(10)}):
\begin{equation}
\beta=\pm\sqrt{\frac{\alpha(1-\alpha)}{N(N+1)}}\,.
\label{(14)}
\end{equation}

Apparently, the character of the solution will depend on the charges $Q$ and $q$ and the
curvature of the internal space, $k$. We examine simple cases in which we can obtain exact
solutions with ansatze (\ref{(13)}).

$\bullet$ Case (i). $Q=0$ and $k>0$. We have the following exact solution:
\begin{equation}
\Delta(r)=\frac{(N-1)^2 k}{4b^2}r^2-\frac{M}{\gamma}\,,
\quad \phi=0\,,\quad R(r)=\gamma r\,,
\label{(15)}
\end{equation}
where $M$ is an integration constant and the compactification scale $b$ is determined by
\begin{equation}
b=\left(\frac{2(N-1)!}{N^2-1}\kappa^2q^2\right)^{1/2(N-1)}\,.
\label{(16)}
\end{equation}

In this case, $\alpha=1$, and then $\beta=0$ in the expression (\ref{(14)}). This solution
is just the BTZ black hole solution \cite{1} with a constant internal space, when
$\gamma=1$. It can be seen that the neutral, rotating BTZ black hole solution is also
admitted for the solution in this case.

Thus the solution has a horizon located at
\begin{equation}
r_+=\frac{4}{N-1}\sqrt{\frac{M}{k\gamma}}b\,.
\label{(17)}
\end{equation}

Note that the horizon length is proportional to the scale of the internal space. The
temperature of the black hole is given by
\begin{equation}
T_H=\frac{1}{4\pi}\left.\frac{d\Delta(r)}{dr}\right|_{r=r_+}=\frac{N-1}{4\pi b}
\sqrt{\frac{Mk}{\gamma}}\,.
\label{(18)}
\end{equation}

$\bullet$ Case (ii). $q=0$ and $k>0$. In this case, we obtain a solution corresponding to
$\alpha=-\beta=1/(N^2+N+1)$. Then the exact solution can be written as
\begin{eqnarray}
&
&\Delta(r)=\frac{(N^2+N+1)^2}{N^2}\frac{k}{b^2}\left(\frac{r}{r_0}\right)^{-2(N+1)\alpha}r^2
-\frac{M}{\alpha\gamma}r^{1-\alpha}\,,\label{(19)}\\ & &\phi(r)=
-\alpha\ln\left(\frac{r}{r_0}\right)\,,\quad R(r)=\gamma r^{\alpha}\,,\label{(20)}
\end{eqnarray}
where $\alpha=1/(N^2+N+1)$, $M$ is an integration constant. There is a relation among
$\gamma$, $b$, $r_0$, $k$ and $Q$:
\begin{equation}
\frac{(N+1)(N-1)^2}{N}\frac{k\gamma^2r_0^{2\alpha}}{b^2}
=\kappa^2Q^2\,.
\label{(21)}
\end{equation}

If $M>0$, the spacetime described by (\ref{(19)}) and (\ref{(20)}) has an event horizon. The
location of the horizon is
\begin{equation}
r_+=\left(\frac{N^2r_0^{-2(N+1)\alpha}}{N^2+N+1}
\frac{b^2M}{k\gamma}\right)^{(N^2+N+1)/N(N-1)}\,.
\label{(22)}
\end{equation}

The temperature of the black hole is given by
\begin{equation}
T_H=\frac{1}{4\pi}\left.
\frac{d\Delta(r)}{dr}\right|_{r=r_+}
=\frac{N(N-1)}{4\pi}\frac{M}{\gamma}\left(
\frac{N^2r_0^{-2(N+1)\alpha}}{N^2+N+1}\frac{b^2M}{k\gamma}
\right)^{-1/N(N-1)}\,.
\label{(23)}
\end{equation}

$\bullet$ Case (iii). $k>0$. In this case we find a solution corresponding to
$\alpha=N/(2N+1)$ and $\beta=-1/(2N+1)$. Then the exact solution can be seen to be
\begin{eqnarray}
& &\Delta(r)=(2N+1)^2\frac{k}{b^2}\left(\frac{r}{r_0}\right)^{2(N+1)\beta}r^2\nonumber \\
& &\quad+
\frac{(2N+1)^2}{(N-1)^2}\frac{\kappa^2Q^2}{R(r)^2}
\left(\frac{r}{r_0}\right)^{2N\beta}r^2-\frac{M}{\alpha\gamma}r^{1-\alpha}\,,\label{(24)}\\
& &\phi(r)=\beta\ln\left(\frac{r}{r_0}\right)\,,\quad R(r)=\gamma r^{\alpha}\,,
\label{(25)}
\end{eqnarray}
where $\alpha=N/(2N+1)$, $\beta=-1/(2N+1)$, and $M$ is an integration constant. The relation
among $\gamma$, $b$, $r_0$, $q$, and $Q$ is
\begin{equation}
(N-1)(N-1)!\frac{\gamma^2q^2r_0^{-2N\beta}}{b^{2N}}
=Q^2\,.
\label{(26)}
\end{equation}

This spacetime described by (\ref{(24)}) and (\ref{(25)}) may also have an event horizon.
The location of the horizon is determined by the equation
\begin{equation}
(2N+1)^2\frac{k}{b^2}\left(\frac{r}{r_0}\right)^{2(N+1)\beta}r^2+
\frac{(2N+1)^2}{(N-1)^2}\frac{\kappa^2Q^2}{R(r)^2}
\left(\frac{r}{r_0}\right)^{2N\beta}r^2-\frac{M}{\alpha\gamma}r^{1-\alpha}=0\,.
\label{(27)}
\end{equation}

For general values of $Q$ there can be one or two roots for (\ref{(27)}) if $M>0$. Therefore
the internal structure of the horizon will be worth studying in general cases. We leave this
subject for future work. We focus on the special case $Q=q=0$. In this case the solution
becomes
\begin{equation}
\Delta(r)=(2N+1)^2\frac{k}{b^2}\left(\frac{r}{r_0}\right)^{2(N+1)\beta}r^2
-\frac{M}{\alpha\gamma}r^{1-\alpha}
\quad (\mbox{when }Q=q=0)\,.
\label{(28)}
\end{equation}
with $\alpha=N/(2N+1)$ and $\beta=-1/(2N+1)$. The horizon is located at
\begin{equation}
r_+=\left(\frac{r_0^{2(N+1)\beta}}{N(2N+1)}
\frac{b^2M}{k\gamma}\right)^{(2N+1)/(N-1)}
\label{(29)}
\end{equation}
and the temperature of the black hole is
\begin{equation}
T_H=\frac{1}{4\pi}\left.
\frac{d\Delta(r)}{dr}\right|_{r=r_+}
=\frac{N(N-1)}{4\pi}\frac{M}{\gamma}\left(
\frac{r_0^{2(N+1)\beta}}{N(2N+1)}\frac{b^2M}{k\gamma}
\right)^{-N/(N-1)}\,.
\label{(30)}
\end{equation}

$\bullet$ Case (iv). $q=0$ and $k=0$. In this case we obtain a solution corresponding to
$\alpha=(N+1)/(2N+1)$ and $\beta=-1/(2N+1)$. The exact solution can be written as
\begin{eqnarray}
& &\Delta(r)=\frac{(2N+1)^2}{N(N+1)}
\frac{\kappa^2Q^2}{\gamma^2r_0^{2N\beta}}-
\frac{M}{\alpha\gamma}r^{1-\alpha}\,,\label{(31)}\\
& &\phi(r)=\beta\ln\left(\frac{r}{r_0}\right)\,,
\quad R(r)=\gamma r^\alpha\,,
\label{(32)}
\end{eqnarray}
where $\alpha=(N+1)/(2N+1)$, $\beta=-1/(2N+1)$, and $M$ is an integration constant. The
spacetime described by (\ref{(31)}) and (\ref{(32)}) has a `cosmological' horizon, i.e.
$\Delta(r)=0$ for $r=r_H$ and $\Delta(r)>0$ for $r<r_H$, where
\begin{equation}
r_H=\left(\frac{2N+1}{N}\frac{\kappa^2Q^2r_0^{-2N\beta}}{\gamma M}
\right)^{(2N+1)/N}\,.
\label{(33)}
\end{equation}

This case yields a singularity at the origin. Thus the solution does not describe a black
hole and is just a solitonic solution.

$\bullet$ Case (v). $Q=0$ and $k=0$. We obtain a solution corresponding to
$\alpha=(N+1)/(5N+1)$ and $\beta=-2/(5N+1)$ in this case. The exact solution is
\begin{eqnarray}
& &\Delta(r)=\frac{(5N+1)^2(N-1)!}{2(N^2-1)}\frac{\kappa^2q^2}{b^{2N}}
\left(\frac{r}{r_0}\right)^{4N\beta}r^2-\frac{M}{\alpha\gamma}r^{1-\alpha}\,,\label{(34)}\\
& &\phi(r)=\beta\ln\left(\frac{r}{r_0}\right)\,,\quad
R(r)=\gamma r^{\alpha}\,,
\label{(35)}
\end{eqnarray}
where $\alpha=(N+1)/(5N+1)$, $\beta=-2/(5N+1)$, and $M$ is an integration constant. This
spacetime described by (\ref{(34)}) and (\ref{(35)}) also has a `cosmological' horizon
located at
\begin{equation}
r_H=\left(\frac{(5N+1)(N - 1)!}{2(N-1)}\frac{\gamma\kappa^2q^2r_0^{-4N\beta}}{b^{2N}M}
\right)^{(5N+1)/2(N-1)}\,.
\label{(36)}
\end{equation}

In this spacetime a singularity lies at the origin. Thus the solution is merely a solitonic
solution.

$\bullet$ Case (vi). $Q=q=0$ and $k=0$. In this case a general solution is
\begin{eqnarray}
& &\Delta(r)=c r^{1-\alpha}\,,\label{(37)}\\
& &\phi(r)=\beta\ln\left(\frac{r}{r_0}\right)\,,\quad
R(r)=\gamma r^{\alpha}\,,\label{(38)}
\end{eqnarray}
where
\begin{equation}
\beta=\pm\sqrt{\frac{\alpha(1-\alpha)}{N(N+1)}}
\label{(39)}
\end{equation}
and $C$ is an integration constant. Obviously this spacetime described by (\ref{(37)}) and
(\ref{(38)}) is singular. For $\alpha=1$, the dilaton field is decoupled and the solution
corresponds to a point singularity with a deficit angle at the origin.

\bigskip

In this paper we have obtained exact solutions to Einstein's equations coupled with
antisymmetric tensor fields in $(2+N+1)$-dimensional spacetime. The solutions describe a
product of a three-dimensional radially symmetric spacetime and an internal space of which
the scale may depend on the radial distance from the origin. Our solutions can be applied
to various supergravity models; for example, $N=2$ supergravity in five dimensions \cite{8}
includes a $U(1)$ gauge field, and then compactification to three dimensions naturally occurs
in such a model.

Classical stability of the solutions will be one of the topics to study in future work. The
evaporation process of black holes of the present type should also be investigated; the extra
dimensions may affect the property of the black hole.

If there has been a `dimensional' phase transition in the very early history of the universe,
the spacetime described by our solutions may arise as topological defects or dislocations.
To determine whether these `wrong' reductions might be able to take place and have an
influence on cosmology we must be better acquainted with the quantum theory of gravity.

We can generalize our work to the case with a dilaton field, which is typically introduced
by string theory. The $3$-form tensor fields strength will naturally induce the compactified
geometry. We plan to discuss these subjects elsewhere.

\section*{Acknowledgements}
We thank K. C. K. Chan and R. B. Mann for important correspondence.



\begin{thebibliography}{99}
\bibitem{1} M.~Ba\~nados, C.~Teitelboim and J.~Zanelli,
Phys. Rev. Lett. \textbf{69} (1992) 1849.

M.~Ba\~nados, M.~Henneaux, C.~Teitelboim and J.~Zanelli,
Phys. Rev. \textbf{D48} (1993) 1506; 
\textbf{D88} (2013) 069902(E).

\bibitem{2} S.~F.~Ross and R.~B.~Mann,
Phys. Rev. \textbf{D47} (1993) 3319.

D.~Cangemi M.~Leblanc and R.~B~Mann, 
Phys. Rev. \textbf{D48} (1993) 3606.

A.~Achucarro and M.~Ortiz, 
Phys. Rev. \textbf{D48} (1993) 3600.

G.~T.~Horowitz and D.~L.~Welch,
Phys. Rev. Lett. \textbf{71} (1993) 328. 

N.~Kaloper,
Phys. Rev. \textbf{D48} (1993) 2598.

C.~Farina, J.~Gamboa and A.~J.~Segui-Santonja,
Class. Quant. Grav. \textbf{10} (1993) L193.

K.~Shiraishi and T.~Maki, 
Class. Quant. Grav. \textbf{11} (1994) 695; 
Phys. Rev. \textbf{D49} (1994) 5286.

A.~R.~Steif, 
Phys. Rev. \textbf{D49} (1994) R585.

G.~Lifschytz and M.~Ortiz, 
Phys. Rev. \textbf{D49} (1994) 1929.

N.~Cruz, C.~Martinez and L.~Pena,
Class. Quant. Grav. \textbf{11} (1994) 2731.

\bibitem{3}
K.~C.~K.~Chan and R.~B.~Mann,
Phys. Rev. \textbf{D50} (1994) 6385; 
\textbf{D52} (1995) 2600(E).


\bibitem{4}
G.~W.~Gibbons and K.~Maeda, 
Nucl. Phys. \textbf{B298} (1988) 741.

\bibitem{5} 
T.~Appelquist, A.~Chodos and P.~G.~O.~Freund (ed),
\textit{Modern Kaluza-Klein Theories}
(Reading, MA: Addison-Wesley).

A.~Salam and E.~Sezgin (ed),
\textit{Supergravities in Diverse Dimensions} vol 2
(Singapore: World Scientific).

M.~J.~Duff, B.~E.~W.~Nilsson and N.~Pope, 
Phys. Rep. \textbf{130} (1986) 1.

V.~M.~Emel'yanov  et al.,
Phys. Rep. \textbf{143} (1986) 1.

D.~Bailin and A.~Love, 
Rep. Prog. Phys. \textbf{50} (1987) 1087.

J.~Strathdee,
Int. J. Mod. Phys. \textbf{A1} (1986) 1.

\bibitem{6}
P.~van Baal, F.~A.~Bais and P.~van Nieuwenhuizen, 
Nucl. Phys. \textbf{B233} (1984) 477.

\bibitem{7} P.~G.~O.~Freund and M.~A.~Rubin, 
Phys. Lett. \textbf{B97} (1980) 233.

\bibitem{8} 
E.~Cremmer,
in \textit{Superspace and Supergravity} ed S.~W.~Hawking and M.~Rocek 
(Cambridge: Cambridge University Press) p. 267.

L.~Dolan, 
Phys. Rev. \textbf{D30} (1984) 2474. 

\end{thebibliography}
\end{document}